# A scalable and inexpensive method to build physically un-clonable functions based on single-walled carbon nanotubes

*Enrique Burzurí\*, Daniel Granados\* and Emilio M. Pérez*

IMDEA Nanoscience, Ciudad Universitaria de Cantoblanco, c\Faraday 9, 28049 Madrid, Spain

*A physically un-clonable function (PUF) is a physical system that cannot be reproduced or predicted and therefore is a good basis to build security and anti-counterfeiting applications. The unclonability of PUFs typically stems from the randmoness induced in a system during sophisticated fabrication methods. It is precisely this built-in complexity the bottleneck hindering scalability and increasing costs. Here, we produce in a simple manner PUFs based on arrays of carbon nanotubes junctions simultaneously assembled by dielectrophoresis. We demonstrate that the intrinsic inhomogeneity of carbon nanotubes at the nanoscale, combined with the unpredictability introduced by liquid phase-based fabrication methods results in unique electronic profiles of easily scalable devices. This approach could be extrapolated to generate PUFs based on other nanoscale materials.*

The growing internet of things (IoT) industry and the needs of the anti-counterfeiting market are nowadays boosting an increasing demand for advanced identification systems compatible with high-security standards[1]. The unclonability and unpredictability are cornerstone and are precisely the core properties of an emerging family of physical systems: the physically unclonable functions (PUFs). A PUF is a measurable quantity that is easy to read and very hard to predict, embedded in a device that is easy to make and almost impossible to duplicate, even if all the details about the manufacturing process are known[2-4]. The main ingredient to fabricate PUFs is essentially randomness[1, 5-8]. This can be intrinsic to the structure or properties of a system or introduced *via* a non-deterministic fabrication process. An example of this fabrication-induced randomness in PUFs is based on polymeric particles where randomness stems from randomly generated silica film wrinkles[1]. Other recent ones use magnetic domains[5] where the variability is induced by the random distribution of magnetization axes during fabrication, among other examples[6-8]. Nanomaterials, like low-dimensional materials integrated in nanodevices offer an interesting alternative for PUFs[9-10]. Small intrinsic variations at the nanoscale result in non-deterministic distinct properties of the manufactured devices at the macroscale. For example, CVD grown $MoS_2$ results in a random distribution of island thicknesses with different optical properties[9].

Intrinsic randomness in structural properties and small variations at the nanoscale are precisely characteristics inherent to commercial single-walled carbon nanotubes (SWNTs) and devices fabricated with them. SWNTs are synthesized as a mixture of different chiralities that directly determine their electronic properties. They can be metallic, quasi-metallic or semiconductors with different well-defined finite bandgap depending on their chirality[11]. In addition, SWNTs present defects (O-containing functionalities, C vacancies, 5 and 7-membered rings, etc.) at the atomic level, which make even two SWNTs with identical chirality different (see Figure 1a). These characteristics, currently seen as the main limiting factor preventing SWNTs to reach real-life marketable applications in nanoelectronics, could be in turn used as an advantage to fabricate PUFs.

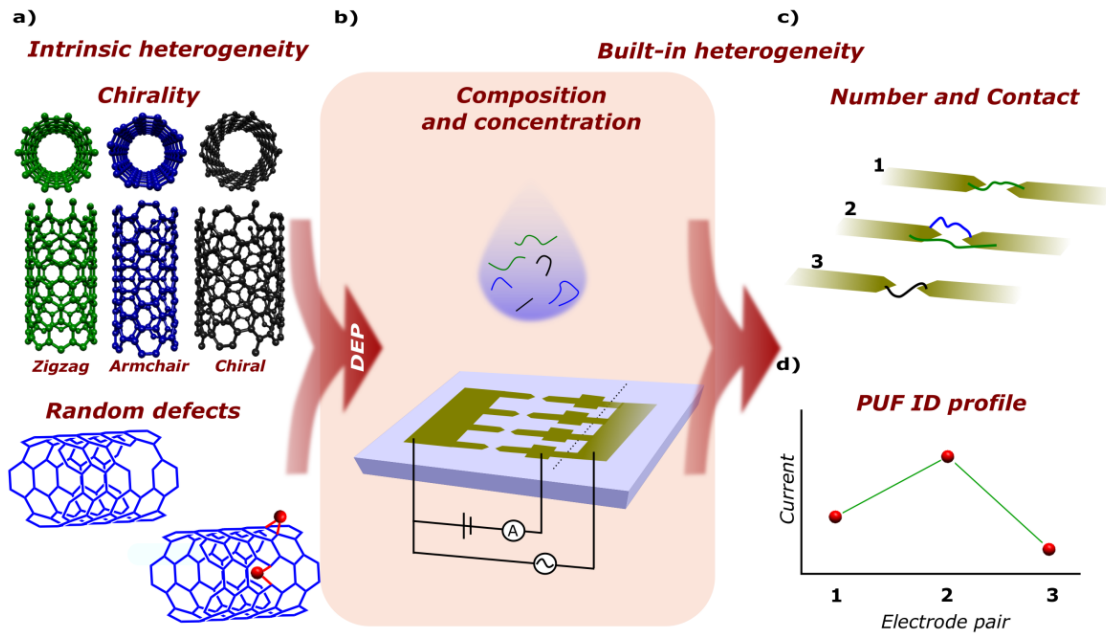

*Figure 1. PUFs made with SWNTs junctions. (a) Intrinsic heterogeneity in commercial SWNTs originates in their different chiralities that derives in different electronic behavior (metallic, quasi metallic or semiconductor with different bandgaps) and random defects in the lattice like vacancies and oxygen functionalities. (b) The SWNTs in the liquid phase are simultaneously transported and positioned by dielectrophoresis in an array of electrodes. The composition of the droplet is heterogeneous in nature and concentration of SWNTs. Each electrode pair is isolated by removing the contact to the common pad (dotted line). (c) The fabrication by DEP results in a randomized number of SWNTs and electrical coupling per electrode. (d) The combination of all these intrinsic and extrinsic variables results in differentiated sets of electrodes with unique current profiles even under the same preparation conditions.*

A second level of randomness is introduced in the fabrication of electronic devices based on SWNT suspensions[12-14]. This is the basis of a pioneering report on CNT-based PUFs where the SWNTs are integrated into a metallic mesh[10]. The inhomogeneous dynamics of the suspension randomizes the distribution of SWNTs on the surface and therefore which cells of the mesh will be connected. This approach is limited to a binary or ternary response of the PUF (empty or full cell) and relies on complex fabrication techniques requiring several nanolithography steps and complex chemical processing.

Here we report the fast and inexpensive fabrication of PUFs based on simple arrays of SWNT junctions fabricated in a single lithography step. The arrays contain 16 parallel pairs of electrodes where the nanotubes are trapped simultaneously from the liquid phase by dielectrophoresis (DEP) (see Figure 1b). This number of electrodes is arbitrary as a proof-of-concept and can be easily scalable to larger or multiple devices. The inhomogeneous composition and dynamics of SWNT suspensions makes it impossible to anticipate the distribution of chiralities present in suspension and if a particular pair of electrodes will be bridged by a single, two or a bundle of SWNTs (see Figure 1c). We show that this simple non-deterministic fabrication approach, combined with the intrinsic heterogeneity of SWNTs, leads to a unique and irreproducible distribution of SWNTs in the device resulting in unique conductance profiles (see Figure 1d) in easy-to-do electron transport measurements. We find Hamming distances compatible with PUFs functionalities. The devices are irreversibly modified under inspection by electron microscopy hindering counterfeiting attempts.

The multi-electrode array consists of a set of parallel submicrometer-spaced Au electrodes connected to two common Au contact pads. One side of the device also has intermediate individualized pads between electrodes and the common contact. The whole device is built on a highly-doped silicon substrate coated with a thin insulating $SiO_2$ layer. The electrodes and pads are fabricated *via* laser mask-less optical lithography and subsequent thermal evaporation of Cr/Au. A scheme of the final device is shown in Figure 1b. A closer view of some details of the devices is shown in the optical images in Figures 2a and 2b.

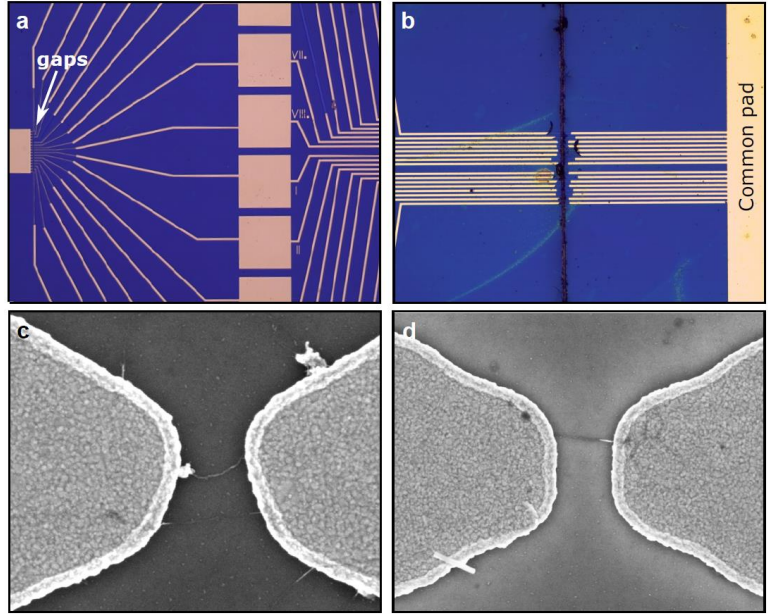

*Figure 2*. *PUF nano-device architecture. (a) (b) Optical images of a multi-electrode device showing: (a) the intermediate individualized pads and (b) the cutting between intermediate and common pads after DEP. (c) (d) Scanning electron microscopy (SEM) images of two electrode pairs of the same device after DEP with two and one SWNT respectively.*

The carbon nanotubes are simultaneously transported and positioned in the multi-electrode array *via* a dielectrophoresis (DEP) mechanism. Dielectrophoresis consists in the directed motion of polarizable particles (like SWNTs) properly dispersed in a liquid medium and under the influence of a non-homogenous electrical field. This technique has shown its potential to achieve fast submicrometer precision in the positioning of SWNTs[12, 15-16] and other nanostructures[17-21] in complex devices. The electrical field in the device is created by applying an alternating bias voltage between the common pads. The dielectrophoretic force is proportional to the electrical field gradient that, in the device geometry, is maximum and directed towards the inter-electrode spacing. The SWNTs are attracted, aligned and trapped between the electrodes.

To maximize the SWNTs heterogeneity in the dispersion, we use commercially available samples of high-pressure carbon monoxide (HiPco)-grown SWNTs. HiPco SWNTs typically contain 13 major (n,m) chiralities, together with at least another 20 other (n,m) species in minor quantities[22-24]. The SWNT suspensions are prepared by direct ultrasonication in organic solvents with polarizabilities lower than SWNTs (See Supporting Information). Chloroform or tetrachloroethane meet this requirement and are good dispersing media for SWNTs. The average degree of individualization of the SWNTs in suspension can be controlled through variation in the ultrasonication time, temperature and power but the exact size, shape and composition of each particle in the suspension is stochastic, therefore, unlike solutions of pure molecular compounds, there is no means to reproduce the exact composition of a SWNT suspension. In practice, a micro-droplet containing the SWNTs suspension is drop-casted onto the device surface while a DEP ac voltage is kept between the electrodes. The design of the device allows carrying out DEP simultaneously in all pairs of electrodes by establishing electrical contact to the common Au pads. Typical DEP parameter values that optimize the probability of trapping a single or few SWNTs between electrodes separated by ca. 500 nm are $V_{ac}$ = 3 V, $v$ = 3 MHz, $t$ =1 min[12]. After DEP, the device is thoroughly rinsed with the same solvent and dried with nitrogen gas to remove excess material. Thereafter, the connections between the common pad and the electrodes are removed, so that each electrode pair can be electrically probed independently. The removal of the connections is achieved by dicing the Au strips with a diamond tip (see Figure 2b) or, alternatively, the substrate area under the connecting strips can be pre-prepared with an adhesive strip that is removed after DEP.

Figure 2(c,d) shows two scanning electron microscopy (SEM) images of two representative electrode pairs in a device containing two and one SWNTs respectively after DEP positioning. The stochasticity of the dielectrophoretic process, together with the composition of each droplet results in a randomized and irreproducible array of electrodes in terms of the number of tubes per electrode and coupling to the electrodes. We observe electrode pairs containing none, one, two

and less frequently, bundles or multiple SWNTs. In addition, the intrinsic SWNTs heterogeneity is also expected to give a randomized composition in chirality and defects per electrode pair. The result is expected to be a unique electrical fingerprint for each device that forms the basis of the SWNT-based PUFs. We note that the non-deterministic nature of the final device is not due to current limitations in the control of liquid-phase-based and dielectrophoresis techniques but a combination of intrinsic randomness in the nanoscale character of the material and technique as described before.

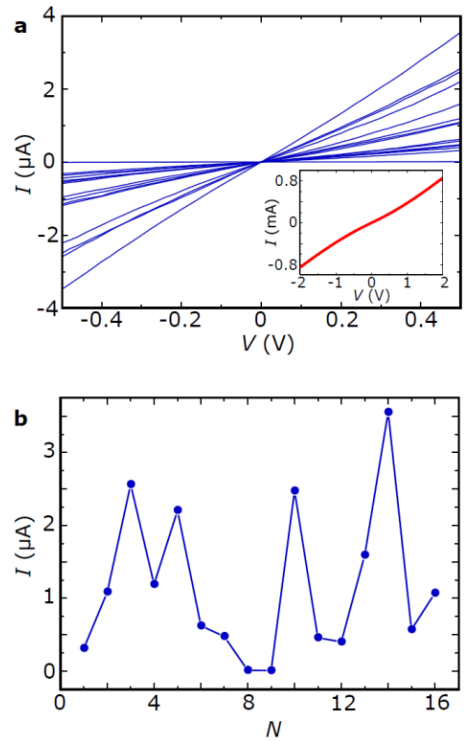

*Figure 3.* *Electron transport characterization of PUFs. (a) IV characteristics of the 16 electrode pairs contained in PUF1 after dicing the connection to the common pad. The shape and absolute value vary randomly from electrode to electrode pair. The inset shows the IV characteristic of the full device before dicing the common connection. (b) Current profile of PUF1 built by taking the I value at V = 0.5 V for each electrode pair in the same device.*

The inset in Figure 3a shows the current $I$ measured as a function of the bias voltage $V$ applied between the common Au terminals, i.e. the collective current of the device, after DEP. The s-shape of the $IV$ indicates the presence of non-ohmic contacts between SWNTs and electrodes. Figure 3a shows the $IV$ curves measured for each individual pair of electrodes after removal of the connection to the common pads. The $IV$ curves clearly differ for each electrode pair in their relative shape and in the absolute current values. The variation is randomly distributed, independently of the electrodes' position in the device. These fluctuations in the conductance are most likely originated by a combination of different tube chirality (and therefore electronic behavior), electronic coupling with the electrodes and number of SWNTs per electrode pair. To define an identifier (ID) that univocally characterizes the PUF we select the current $I$ measured at a fixed $V = 0.5$ V and represent it as a function of the electrode pair number $N$. The resulting current profile or PUF ID is shown in Figure 3b.

The variability in this analogic ID for each PUF can span orders of magnitude, as seen in additional profiles in the Supporting Information. While this is *per se* an advantage for these PUF architecture, in contrast with previous binary CNT-based PUFs[10], it can complicate the direct comparison between different PUF IDs. The raw $I$ profile is therefore translated or normalized for practical purposes into a set of quantized bits $I^N$ that can be easily compared from device to device. The highest conductance value is assigned a 16, *i.e.* the maximum number of electrodes. The rest are assigned a decreasing integer value for decreasing conductance down to 1 for the lowest conductive set. The resulting equivalent $I^N$ profile is shown in Figure 4a. A comparison with another PUF ID profile (PUF2) is shown in Figure 4b. See Supporting Information for additional profiles. The raw Hamming distance between PUF1 and PUF2, defined as the number of changes needed to make two strings equal, is $H_{raw}$ (1-2) = 15 in this case, i.e.: only one bit is coincident.

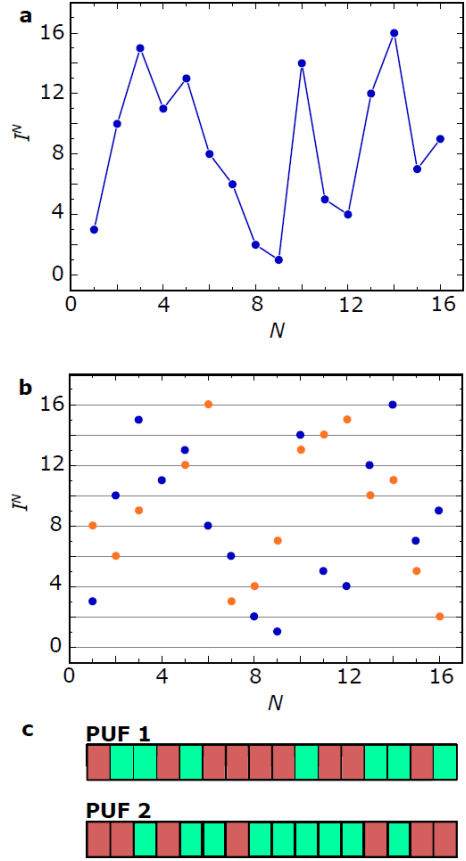

*Figure 4.* *PUF ID construction. (a) Normalized profile $I^N$ for PUF1. The highest conductance value is assigned a $I^N$ =16 bit (the maximum number of electrodes). The rest of I values are assigned a descending integer bit for descending current value down to 1 for the lowest I. (b) Comparison of $I^N$ profiles of PUF1 (blue) and PUF2 (orange). Only the N = 4 value is coincident. The raw Hamming distance is therefore $H_{raw}$(1-2) = 15. (c) Binary equivalent profiles for PUF1 and PUF2. Red = 0, green = 1. The bit N in the position i+1 is set to 0(1) when $I^N$ follows $I^N_{i+1}<I^N_i$($I^N_{i+1}>I^N_i$). The initial bit $N_1$ is set to 0 for $I^N_1 \leq 8$ or to 1 otherwise.*

The devices comprise 16 pairs of electrodes with 16 corresponding non-correlated $I$ levels. The number of possible different ID configurations is therefore $16^{16}$ = 1.84x10$^{19}$. Accordingly, the probability of repeating or cloning one normalized configuration would be 1/1.84x10$^{19}$ = 5.4x10$^-$ [20]. This value can still be easily scaled up by changing the number of electrodes per device or keeping the raw $I$ value for nearly continuous spectra of profiles. To further compare the uniqueness and stability of the profiles we have further simplified the PUF ID to a binary system and calculated the normalized binary Hamming inter- and intra-distances (see Supporting Information for details). The bit $N$ in the position i+1 is set to 0(1) when $I^N$ follows $I^N_{i+1}<I^N_i (I^N_{i+1}>I^N_i)$. This condition can be easily measured in the device. The initial bit $N_1$ is set to 0 for $I^N_1 \leq 8$ or to 1 otherwise. The binary IDs for PUF1 and PUF2, graphically represented in Figure 4c, are: PUF1 ID: 0110100001001101 and PUF2 ID: 0010110111110100. The resulting normalized binary Hamming inter-distance $H^{bin}_{inter}$ between the profiles PUF1 and is $H^{bin}_{inter}$(1-2) = 0.5. By repeating this process to the full set of measured PUFs, the mean Hamming inter-distance $H^{mean}_{inter}$ of the set can be calculated (see Supporting Information). We obtain $H^{mean}_{inter}$ = 0.51. A value close to 0.5 is signature of a random and uncorrelated set of binary PUFs, key to achieve high degrees of un-clonability.[5, 9] On the other hand, the normalized Hamming intra-distance, that measures the divergence of a same profile between two measurements, is typically found below 0.1 (see Supporting Information). In other words, the response of the PUF to a same challenge is reproducible in time. This reproducibility could be further protected by encapsulation of the PUF in a dielectric medium.

In conclusion, we have taken advantage of the intrinsic inhomogeneity at the nanoscale of commercial SWNTs and combine it with simple non-deterministic fabrication methods to fabricate scalable PUFs. We show that two devices fabricated under the same conditions will present different conductivity profiles with extremely low probabilities of repetition. Our study opens the door to the incorporation of other low-dimensional materials like nano-rods, liquid-phase exfoliated 2D materials and heterostructures as PUF identifiers.

## ASSOCIATED CONTENT

**Supporting Information**.

The following files are available free of charge.

Supporting Information file containing additional details about dielectrophoresis, additional PUF profiles and details of the calculations of the Hamming intra- and inter-distances.

## AUTHOR INFORMATION

**Corresponding Author**

* E-mail: Enrique.burzuri@imdea.org
* E-mail: Daniel.granados@imdea.org

**Notes**

The authors declare no competing financial interest.

## ACKNOWLEDGMENT

E.B. thanks funds from the MSCA-IF European Commission programme (grant 746579) and Programa de Atracción del Talento Investigador de la Comunidad de Madrid (2017-T1/IND-5562). E.M.P. acknowledges funding for the European Research Council (ERC-StG-MINT 307609), the Ministerio de Economía y Competitividad (CTQ2014-60541-P, CTQ2017-86060-P), and the Comunidad de Madrid (MAD2D-CM program S2013/MIT-3007). This work is partially supported by the Spanish Ministry of Economy, Industry and Competitiveness through Grant SUPERMNAN ESP2015-65597-C4-3-R DETECTA ESP2017-86582-C4-3-R. D.G.

acknowledges Grant RYC-2012-09864. IMDEA Nanociencia acknowledges support from the 'Severo Ochoa' Programme for Centres of Excellence in R&D (MINECO, Grant SEV-2016-0686).